\documentclass[nohyper,12pt,letterpaper]{JHEP3}

\usepackage[latin1]{inputenc}
\usepackage{bbm,amsfonts}
\usepackage{amssymb,amsmath}

\usepackage{amsfonts}
\usepackage{amsbsy}
\usepackage{concmath, palatino}
\usepackage{mathrsfs}
\usepackage{epic}

\author{CarloAlberto Ratti
\\\\
Dipartimento di Matematica e Fisica E. De Giorgi, Universit\`a del Salento and
INFN, Sezione di Lecce, via per Arnesano 73100 Lecce, Italy 
\\ \\
E-mail: \email{carloalberto.ratti@le.infn.it}}

\abstract{
We consider classical multi-spin string solutions in marginally deformed $AdS_4\times CP^3$.
We reconsider the results of \cite{Wu} and extend them to the non supersymmetric 3-deformation parameter background.
After a general discussion, we study various explicit solutions and we give expressions for their conserved classical angular momenta and energy.
}

\preprint{}

\title{Notes on Multi-Spin Strings in $AdS_4\times CP^3$ and its marginal deformations}


\csname @addtoreset\endcsname{equation}{section}

\def\bseq{\begin{subequation}}  
\def\eseq{\end{subequation}}
\def\bsea{\begin{subeqnarray}}  
\def\esea{\end{subeqnarray}}

\newcommand{\beq}{\begin{equation}}
\newcommand{\bea}{\begin{eqnarray}}
\newcommand{\eea}{\end{eqnarray}}
\newcommand{\eeq}{\end{equation}}

\renewcommand{\a}{\alpha}
\renewcommand{\b}{\beta}

\def\Tr{\textstyle{Tr}}

\begin{document}

\section{Introduction}

Dualities between string theories and gauge theories have been one of the most attractive ideas for the the theoretical community for decades.
After the conjecture of an AdS/CFT duality \cite{AdS/CFT}, a lot of work has been done to produce concrete examples and proofs of it.

Recently a new example of the duality has been proposed \cite{ABJM}, which relates type IIA superstring theory in $AdS_4 \times CP^3$ and ABJM theory.
This is a 3-dimensional superconformal $\mathcal{N}=6$ Chern-Simons theory with gauge group $SU(N)_{k} \times SU(N)_{-k}$ with matter superfields in the bi-fundamental representation of the gauge group.
A lot of efforts has been done to understand the precise correspondence between string solutions in this background and the corresponding gauge-invariant operators in the dual gauge theory. 
For a comprehensive review and a complete list of references, we refer the reader to \cite{review}. 


ABJM is a theory with a large amount of supersymmetry.
It is of general interest to investigate examples of less supersymmetric gauge/string dualities.
This aspect assumes a crucial importance if we are interested in modeling realistic physics.
On the gauge theory side marginal deformations \cite{LS} are a method to partially or completely break the supersymmetry in a controlled way. 
On the string theory side, in \cite{TsT1}, \cite{TsT2} an intriguing solution-generating technique, called the TsT transformation, has been presented.
This technique allows one to build up stringy backgrounds duals to marginal deformed CFT's directly from the undeformed ones.

TsT transformations have been applied to the $AdS_4 \times CP^3$ background in \cite{Imeroni:2008cr}.
In that paper it was pointed out that there are three possible independent marginal deformations of the ABJM theory, each of which controlled by one deformation parameter $\gamma_i$, {\footnotesize  $(i=1,2,3)$}.
If all but one deformation parameter are null, the deformed background still preserves four supercharges and it is dual to the $\beta$-deformed ABJM gauge theory.
In all other cases all the supersymmetries are broken.

Strings in marginally deformed $AdS_4 \times CP^3$ have already been considered in literature. 
In \cite{giant} giant magnons and single spike solutions in $\beta$-deformed $AdS_4 \times CP^3$ were studied.
The finite-size effect on the dispersion relation of giant magnons was found in \cite{giant-fin}.
In \cite{Wu} multi-spin string solutions are considered, both in the undeformed scenario (see also \cite{Wu-Chen}) and in the $\beta$-deformed one.

In this paper we extend the analysis of classical multi-spin string solutions to the non-supersymmetric 3-deformation parameter background.
Before doing so, we reconsider from the beginning the case of multi-spin strings in $\beta$-deformed $AdS_4 \times CP^3$.
We then compare our solutions in the non-supersymmetric background with the supersymmetric ones by setting to zero all but one deformation parameter.

The paper is organized in three main Sections in which we consider classical multi-spin string solutions in undeformed $AdS_4 \times CP^3$ (Section \ref{undeformed}), in supersymmetric $\beta$-deformed $AdS_4 \times CP^3$ (Section \ref{deformed}) and in non-supersymmetric 3-parameter deformed $AdS_4 \times CP^3$ (Section \ref{3deformed}).
In the Appendices we collect details regarding the equations of motion in the deformed backgrounds.

\section{Multi-spin strings in undeformed $AdS_4\times CP^3$}
\label{undeformed}

\subsection{General studies}

In \cite{ABJM} a new $AdS/CFT$ correspondence has been established. 
On the gauge side, the theory involved in the duality is a three-dimensional superconformal theory Chern-Simons-matter theory with
gauge group $SU(N)_k\times SU(N)_{-k}$. The theory has two bifundamental matter superfields $A_1, A_2$
and two anti-bifundamental ones $B_1, B_2$. The superpotential is given by
\begin{equation}
 W=\frac{4\pi}{k}\Tr(A_1B_1A_2B_2-A_1B_2A_2B_1).
 \label{superpotential}
\end{equation}

In \cite{ABJM} it has been established that this theory has a stringy dual description in terms of type IIA string theory in $AdS_4\times CP^3$ geometry.
More precisely the background metric is
\begin{eqnarray}
\label{undef-metric}
 ds^2 &=& R^2(\frac{1}{4}  ds^2_{AdS_4}+ds^2_{CP^3}), 
 \nonumber \\
 ds^2_{AdS_4} &=& -\cosh^2\rho dt^2+d\rho^2+\sinh^2\rho(d\theta^2+\sin^2\theta d\varphi^2), 
 \nonumber \\
 ds^2_{CP^3} &=& d\xi^2+\frac{1}{4} \cos^2\xi(d\theta_1^2+\sin^2\theta_1
 d\varphi_1^2)+\frac{1}{4} \sin^2\xi(d\theta_2^2+\sin^2\theta_2 d\varphi_2^2))\nonumber\\
 &&\,\,\,+\cos^2\xi\sin^2\xi
 (d\psi+\frac{1}{2} \cos\theta_1d\varphi_1-\frac{1}{2} \cos\theta_2d\varphi_2)^2,
\end{eqnarray}
where the range of the angles is
\begin{equation}
 0\le\xi<\frac\pi2, \quad  
 -2\pi\le\psi<2\pi, \quad 
 0\le\theta_i\le\pi, \quad 
 0\le\varphi_i<2\pi.
\end{equation}
The complete background includes also a RR $F_4$ form-flux along the $AdS_4$ directions and RR $F_2$ fluxes in $CP^1\subset CP^3$.
The NS-NS fields are null \cite{ABJM}.

The relation between $R$ and the 't~Hooft coupling $\lambda=N/k$ is
\begin{equation} 
 R=2^{5/4}\pi^{1/2}\lambda^{1/4}\alpha^{\prime1/2}.\label{r}
\end{equation}

We want to study multi-spin solutions moving in this particular background. We start from the general expression for the Polyakov action
\begin{eqnarray}
\label{Polyakov}
 S &=&  \frac{R^2}{4\pi\alpha^\prime}\int d^2 \sigma 
 \bigg( 
 \sqrt{-h} h^{\a\b} \partial_{\a} X^{\mu} \partial_{\b} X^{\nu} g_{\mu\nu} 
 + \varepsilon^{ab} B_{\mu \nu} \partial_a X^{\mu} \partial_b X^{\nu}
 \bigg).
\end{eqnarray}
If we explicit this expression for the background (\ref{undef-metric})
and if we work in the conformal gauge, the action reads:
\begin{eqnarray}
S &=&\frac{R^2}{4\pi\alpha^\prime}\int d\sigma d\tau \left(\frac{1}{4} (\cosh^2\rho(\dot{t}^2-t^{\prime2})
-\dot{\rho}^2+\rho^{\prime2}+\sinh^2\rho(-\dot{\theta}^2+\theta^{\prime2})\right.\nonumber\\
&&\,\,+\sinh^2\rho\sin^2\theta(-\dot{\varphi}^2+\varphi^{\prime 2}))-\dot{\xi^2}+\xi^{\prime2}+\frac{1}{4} \cos^2\xi(\theta_1^{\prime2}+\sin^2\theta_1\varphi^{\prime2}_1\nonumber\\
&&\,\,-(\dot{\theta}_1^2
+\sin^2\theta_1\dot{\varphi}_1^2))+\frac{1}{4} \sin^2\xi(-(\dot{\theta}_2^2+\sin^2\theta_2\dot{\varphi}_2^2)
+\theta_2^{\prime2}+\sin^2\theta_2\varphi^{\prime2}_2)\nonumber\\
&&\,\,+\sin^2\xi\cos^2\xi\left(-(\dot{\psi}+\frac{1}{2} \cos\theta_1\dot{\varphi}_1-\frac{1}{2} \cos\theta_2\dot{\varphi}_2)^2
\right.\nonumber\\ &&\,\,\left.\left.
+(\psi^\prime+\frac{1}{2} \cos\theta_1\varphi_1^\prime-\frac{1}{2} \cos\theta_2\varphi_2^\prime)^2\right) \right).
\end{eqnarray}
From this expression it is immediate to compute the equations of motion for classical strings.
Since we are interested in rotating multi-spin string, we find convenient to use the following Ansatz
\begin{eqnarray}
\label{ansatz}
 && \rho=0, \quad  t=\kappa\tau, \quad \psi=\omega_1\tau+n_1\sigma,
 \nonumber \\
 && \varphi_1=\omega_2\tau+n_2\sigma, \quad  \varphi_2=\omega_3\tau+n_3\sigma, 
\end{eqnarray}
with $\xi, \theta_1, \theta_2$ being constants to be fixed conveniently.   
The periodicity of the angular coordinates $\psi, \varphi_1, \varphi_2$ requires $n_2$ and $n_3$ to be integers while $n_1$ to be an even integer. 

By using (\ref{ansatz}) it follows that the only non-trivial equations of motion are the ones for $\xi, \theta_1$ and $\theta_2$.
More precisely they read
\begin{eqnarray}
\label{xi}
  \frac{1}{2} \sin4\xi(\tilde\omega^2-
\tilde n^2)-\frac{1}{4} \sin2\xi(\sin^2\theta_1(\omega_2^2-n_2^2)+
\sin^2\theta_2(\omega_3^2-n_3^2))=0,
\end{eqnarray}
\begin{equation}
\label{th1}
 (-\omega_2\tilde\omega+n_2\tilde n)\sin\theta_1
 \sin^2\xi\cos^2\xi+\frac{1}{4} \cos^2\xi\sin2\theta_1(\omega_2^2-n_2^2)=0,
\end{equation}
\begin{equation}
\label{th2}
 (\omega_3\tilde\omega-n_3\tilde n)\sin\theta_2
 \sin^2\xi\cos^2\xi+\frac{1}{4} \sin^2\xi\sin2\theta_2(\omega_3^2-n_3^2)=0,
\end{equation}
where we have defined
\begin{eqnarray}
 \tilde\omega&\equiv&\omega_1+\frac{\omega_2}{2}\cos\theta_1-\frac{\omega_3}{2}\cos\theta_2,\\
\tilde n&\equiv& n_1+\frac{n_2}{2}\cos\theta_1-\frac{n_3}{2} \cos\theta_2.
\end{eqnarray}
The Virasoro constraints are
\begin{eqnarray}
\label{vir1}
k^2 &= &
4 \sin^2 \xi\cos^2 \xi(\tilde \omega^2 + \tilde n^2) +\cos^2\xi\sin^2\theta_1(\omega_2^2+n_2^2)\nonumber\\
&&\quad  + \sin^2\xi\sin^2\theta_2(\omega_3^2+n_3^2), 
\\
\label{vir2} 
0 &= & \frac{1}{4} (\omega_2n_2\cos^2\xi\sin^2\theta_1 +\omega_3n_3\sin^2\xi\sin^2\theta_2)+\tilde\omega\tilde n\sin^2\xi\cos^2\xi. 
\end{eqnarray}
The energy and the angular momenta of the string are
\begin{eqnarray}
\label{charges}
E
&=&
\frac{\sqrt{\tilde\lambda}}{4} \kappa,
\\
J_{\psi}
&=&
\sqrt{\tilde\lambda}\int\frac{d\sigma}{2\pi}\tilde\omega\sin^2\xi\cos^2\xi
\nonumber\\
&=&
\sqrt{\tilde\lambda}\tilde\omega\sin^2\xi\cos^2\xi,
\\
J_{\varphi_1}
&=&
\sqrt{\tilde\lambda}(\frac{1}{2} \tilde\omega\sin^2\xi\cos^2\xi\cos\theta_1+\frac{\omega_2}{4} \cos^2\xi\sin^2\theta_1),
\\
J_{\varphi_2}
&=&
\sqrt{\tilde\lambda}(-\frac{1}{2} \tilde\omega\sin^2\xi\cos^2\xi\cos\theta_2+\frac{\omega_3}{4} \sin^2\xi\sin^2\theta_2).
\end{eqnarray}

\subsection{Special solutions}
\label{special-undefo}
\begin{itemize}
\item Solution 1 \\
 Let's consider the following two choices for the 3 embedding coordinates that we have not fixed in (\ref{ansatz}), namely $\xi$, $\theta_1$ and $\theta_2$ 
\begin{eqnarray}
 \textrm{1.1)} && \xi=\pi/4, \ \theta_1=\theta_2=0,
 \label{soluz1.1}
 \\
 \textrm{1.2)} && \xi=\pi/4, \ \theta_1=0,\  \theta_2=\pi.
 \label{soluz1.2}
\end{eqnarray}
In this case, all the equations of motion are satisfied for arbitrary
$\omega_i$, $n_i$, $(i=1, 2, 3)$.
The two Virasoro constraints give respectively
\begin{equation}
 \label{vir11}
 \kappa^2= (\omega_1+\frac{\omega_2}{2}\mp \frac{\omega_3}{2})^2+(n_1+\frac{n_2}{2}\mp \frac{n_3}{2})^2,
\end{equation}
and
\begin{equation}
 \label{Vir12}
 (\omega_1+\frac{\omega_2}{2} \mp \frac{\omega_3}{2})(n_1+\frac{n_2}{2}\mp \frac{n_3}{2})=0.
\end{equation}
Here the minus/plus signs corresponds to the two choices (\ref{soluz1.1})/(\ref{soluz1.2}).

There are two possible solutions of eq. (\ref{Vir12}):
\begin{eqnarray}
\label{soluzioni}
 && n_1+\frac{n_2}{2} \mp \frac{n_3}{2} = 0,
 \\
 && \omega_1+\frac{\omega_2}{2} \mp \frac{\omega_3}{2} = 0.
\label{sol2}
\end{eqnarray}
In the first case we have
\begin{equation}
 J_{\psi}=2J_{\varphi_1}=\mp 2J_{\varphi_2}=\frac{\sqrt{\tilde\lambda}}{4}(\omega_1+\frac{\omega_2}{2}\mp \frac{\omega_3}{2}),
\end{equation}
We can compute the value of the energy from eq. (\ref{vir11}) and (\ref{charges}). The result is
\begin{equation}
 E=\frac{\sqrt{\tilde\lambda}}{4}\kappa=|J_{\psi}|.
\end{equation}
The value of the charges shows that this solution is actually dual to the chiral primary operators of ABJM theory and it has been identified as the ground state of type II A string in the Penrose limit 
\cite{BMNstring}.

For the second solution (\ref{sol2}), we have 
\begin{eqnarray}
 && J_{\psi} = J_{\varphi_1} = J_{\varphi_2} = 0,
 \nonumber \\
 && E=\frac{\sqrt{\tilde\lambda}}{4}\kappa=\frac{\sqrt{\tilde\lambda}}{4}|n_1+\frac{n_2}2\mp \frac{n_3}2|.
\end{eqnarray}
This solution is less interesting for us, since all the angular momenta of this string are null.

\item Solution 2 \\
Let's consider 
this different specific Ansatz
\begin{eqnarray}
\label{case2}
\xi=\pi/4, \quad \theta_2=\pi-\theta_1, \quad  \omega_2=\omega_3,  \quad n_2=n_3.
\end{eqnarray}

In this case the equation of motion (\ref{xi}) for $\xi$ is satisfied.
The equations of motion (\ref{th1}), (\ref{th2}) for $\theta_1$, $\theta_2$  lead to the same condition, namely
\begin{equation}
 \sin\theta_1(\omega_1\omega_2-n_1n_2)=0.
\end{equation}
If we assume that in this case $\theta_1$ is neither $0$ nor $\pi$, we get
\begin{equation}
 \omega_1\omega_2=n_1n_2.\label{omegan}
\end{equation}
Under this condition, the  Virasoro contraints (\ref{vir1}), (\ref{vir2}) read
\begin{eqnarray}
\label{virarira1}
&&  \kappa^2=\omega_1^2+n_1^2+\omega_2^2+n_2^2+2\cos\theta_1(\omega_1\omega_2+n_1n_2)
\nonumber \\
&& \phantom{\kappa^2} = \omega_1^2+n_1^2+\omega_2^2+n_2^2+4\cos\theta_1\omega_1\omega_2,
\\
\label{virarira2}
 && 0 \phantom{a} = (\omega_1n_2+\omega_2n_1)\cos\theta_1+\omega_1n_1+\omega_2 n_2.
\end{eqnarray}
The first constraint fixes $k$ (i.e. the energy of the solution) while the second one links one of the frequencies $\omega_i$ (or the mode numbers $n_i$) to the other free parameters.

The conserved charges relative to this solution are
\begin{eqnarray}
\label{cariche-undefo}
E & = & \frac{\sqrt{\tilde{\lambda}}}{4}\ \kappa=\frac{\sqrt{\tilde\lambda}}{4}\sqrt{\omega_1^2+n_1^2+\omega_2^2+n_2^2+4\cos\theta_1\omega_1\omega_2},
\nonumber \\
J_{\psi} & = & \frac{\sqrt{\tilde\lambda}}{4}(\omega_1+\omega_2\cos\theta_1),
\nonumber 
\\
J_{\varphi_1} = J_{\varphi_2}&=&
\frac{\sqrt{\tilde\lambda}}{8}(\omega_2+\omega_1\cos\theta_1).
\end{eqnarray}

\end{itemize}

\section{Multi-spin strings in $\beta$-deformed $AdS_4\times CP^3$}
\label{deformed}

\subsection{General discussion}
We consider now the easiest possible marginal deformation of the ABJM theory where only a single deformation parameter is introduced. 
This theory is known as $\beta$-deformed ABJM theory and it preserves four supercharges, i.e. it is an $\mathcal{N} = 2$
supersymmetric three-dimensional gauge theory. $\beta$ denotes the deformation parameter and in general is a complex number. 
Nevertheless we will consider only the case of real deformation parameters.
Following the literature, we use the letter $\gamma$ instead of $\beta$ to denote a real deformation parameter.

From the field theory point of view, the marginal deformation acts by promoting each product among the fields in the Lagrangian of the theory to be the $\ast$-product \cite{trentotto}:
\begin{equation}
\label{star}
f g \longrightarrow f \ast g = e^{i \gamma \pi (Q^f_i Q^g_j - Q^f_j Q^g_i )} f g .
\end{equation}
Here the $Q$'s are appropriately chosen charges of the fields $f$, $g$ with respect to two $U(1)$ symmetries of the Lagrangian.

The gravity dual of marginally deformed theories can be computed through TsT transformations. This is a well known solution generating technique \cite{TsT1}, \cite{TsT2}
that efficiently produces deformed backgrounds from undeformed geometries which present at least two $U(1)$ isometries.
The isometries of the metric are the gravity dual version of the $U(1)$ symmetries of the Lagrangian on the gauge side that lead to the $\ast$-product as defined in (\ref{star}).

In more detail, the TsT transformation consists in a sequence of 3 operations on the original metric: 1) a T-duality performed on a $U(1)$ isometry of the undeformed metric; 2) a shift which mixes the coordinates relative to the two $U(1)$ isometries.
The mixing parameter is the deformation parameter; 3) a second T-dualization on the first $U(1)$. Marginally deformed backgrounds are generated  when both the $U(1)$'s involved in the TsT transformation are in the internal space of the undeformed geometry.

Marginally deformed $AdS_4 \times CP^3$ backgrounds were computed for the first time in \cite{Imeroni:2008cr}. 
There it was also pointed out that there are 3 $U(1)$ isometries in $CP^3$.
In the parametrization we are using, they are realized along the angles $\psi$, $\varphi_1$ and $\varphi_2$. 
Thus in principle we can marginally deform the $AdS_4 \times CP^3$ geometry operating with three independent TsT transformations, 
each depending on different deformation parameters $\gamma_i$, {\footnotesize $i = (1,2,3)$}.
If we restrict ourself to perform only one TsT on the $U(1)$ isometries relative to the coordinates $\varphi_1$ and $\varphi_2$ we get a deformed background dual to the $\beta$-deformed ABJM theory. This is the background we are going to consider in this Section. 

Since we are interested in studying classical closed string solutions, we only need the deformed metric and the NS-NS $B$-field. They are \cite{Imeroni:2008cr}
\begin{eqnarray}
 ds^2_{\tilde\gamma}&=&R^2(\frac{1}{4} ds^2_{AdS_4}+ds^2_{CP^3_\gamma}),
\nonumber \\
\nonumber \\
 ds^2_{CP^3_{\tilde\gamma}}
 &=&
 d\xi^2
 +\frac{1}{4} \cos^2\xi(d\theta_1^2+G\sin^2\theta_1d\varphi_1^2)
 +\frac{1}{4} \sin^2\xi(d\theta_2^2 \,\,+G\sin^2\theta_2d\varphi_2^2)
 \nonumber\\
 & &
 +G\cos^2\xi\sin^2\xi(d\psi+\frac{1}{2} \cos\theta_1d\varphi_1 \,\,-\frac{1}{2} \cos\theta_2d\varphi_2)^2
 \nonumber\\
 & &
 +\tilde{\gamma}^2 G \sin^4\xi \cos^4\xi \sin^2\theta_1 \sin^2\theta_2 d\psi^2,
\\
\nonumber \\
 B &=& -\frac{R^2}{2} \cos^2 \xi\sin^2 \xi\tilde\gamma G \Big( \cos^2\xi\sin^2\theta_1\cos\theta_2d\psi\wedge d\varphi_1
 \nonumber \\
 && \qquad \qquad + \sin^2\xi
\sin^2\theta_2\cos\theta_1d\psi \wedge d\varphi_2
+\frac{1}{2}  fd\varphi_1\wedge d\varphi_2 \Big).
\end{eqnarray}
Here we have defined
\begin{eqnarray}
\label{fG}
 && f = \sin^2\theta_1\sin^2\theta_2+\cos^2\xi\sin^2\theta_1\cos^2\theta_2+\sin^2\xi\sin^2\theta_2\cos^2\theta_1,
\nonumber \\
 && \qquad  G = \frac{1}{1+\tilde\gamma^2f\sin^2\xi\cos^2\xi}, \qquad \qquad 
 \tilde\gamma = \frac{R^2}{4}\gamma.
\end{eqnarray}

To avoid proliferations of signs in the complicated formulas that follow, from now on all the $\gamma$ must be considered as $\tilde\gamma$ unless explicitly specified.

The Polyakov action (\ref{Polyakov}) in this background is
\begin{eqnarray}
\label{azione-1defo}
 S &=& \frac{R^2}{16\pi\alpha^\prime}\int d\sigma d\tau
   \Big\{ 
   \nonumber \\
   &&
   -\cosh^2(\rho) \left(t^{\prime 2}-\dot{t}^2\right)
     +\rho^{\prime 2} -\dot{\rho}^2
   +\sinh^2 (\rho) \left(\theta^{\prime 2} -\dot{\theta}^2\right)
   -4 \dot{\xi}^2 +4 \xi^{\prime 2}    
   \nonumber \\
   &&
   +\sin^2(\theta) \sinh^2(\rho) \left(\varphi^{\prime 2} -\dot{\varphi}^2\right)
   +\cos^2(\xi) \left(\theta_1^{\prime 2}- \dot{\theta}_1^2\right)
   +\sin^2(\xi) \left(\theta_2^{\prime 2}   -\dot{\theta}_2^2\right) 
   \nonumber \\
   &&
   +\frac{G}{4} \sin^2 (2 \xi) \left(4 + \gamma^2 \sin^2(\theta_1) \sin^2(\theta_2) \sin^2(2 \xi)\right) \left(\psi^{\prime 2}- \dot{\psi}^2 \right)
   \nonumber \\
   &&
   +G \cos^2(\xi) \left(\cos^2(\theta_1) \sin^2(\xi)+\sin^2(\theta_1)\right) \left(\varphi_1^{\prime 2}-\dot{\varphi}_1^2\right) 
   \nonumber \\
   &&
   +G \sin^2(\xi) \left(\cos^2(\theta_2) \cos^2(\xi)+\sin^2(\theta_2)\right) \left(\varphi_2^{\prime 2} -\dot{\varphi}_2^2\right)
   \nonumber \\
   &&
   +4 G \cos (\theta_1) \sin^2 (\xi) \cos^2 (\xi) \left( \psi^{\prime} \varphi_1^{\prime}-\dot{\psi} \dot{\varphi}_1 \right)
   \nonumber \\
   &&
   -4 G \cos (\theta_2)
   \sin^2(\xi) \cos^2(\xi) \left(\psi^{\prime} \varphi_2^{\prime} -\dot{\psi} \dot{\varphi}_2 \right)
   \nonumber \\
   &&
   -2 G \cos (\theta_1) \cos (\theta_2) \sin^2(\xi) \cos^2(\xi) \left(\varphi_1^{\prime} \varphi_2^{\prime}-\dot{\varphi}_1 \dot{\varphi}_2 \right)
   \nonumber \\
   &&
   +4 \gamma G \sin^2 (\theta_1) \cos (\theta_2) \sin^2 (\xi) \cos^4(\xi) \left(\psi^{\prime} \dot{\varphi}_1 -\dot{\psi} \varphi_1^{\prime} \right)
   \nonumber \\
   &&
   +4 \gamma G \cos (\theta_1) \sin^2(\theta_2) \sin^4(\xi) \cos^2(\xi) \left(\psi^{\prime} \dot{\varphi}_2 - \dot{\psi} \varphi_2^{\prime} \right)
   \nonumber \\
   &&
   + 2 \gamma f G \sin^2 (\xi) \cos^2(\xi) \left(\varphi_1^{\prime} \dot{\varphi}_2 - \dot{\varphi}_1 \varphi_2^{\prime} \right)
\Big\}.
\end{eqnarray}
The deformation parameter $\gamma$ appears either explicitly or inside the expressions for $G$ and $f$.
From this formula the computation of the equations of motion is straightforward. 

Since we are interested in multi-spin string solutions, we still use the general Ansatz (\ref{ansatz}). We also require that $\xi$, $\theta_1$ and $\theta_2$ are constants.
As in the undeformed case, the equations of motion for $t$, $\rho$, $\theta$, $\varphi$, $\psi$, $\varphi_1$ and $\varphi_2$ are automatically satisfied.

The equations of motion for $\xi$, $\theta_1$ and $\theta_2$ read
\begin{eqnarray}
\label{eomxi}
&&
A_{\xi}^{\gamma}\  G + B^{\gamma}\  \frac{\partial G}{\partial \xi} =0,
\qquad 
\label{eomth1}
A^{\gamma}_{\theta_1}\  G + B^{\gamma}\  \frac{\partial G}{\partial \theta_1} = 0,
\qquad 
A^{\gamma}_{\theta_2}\  G + B^{\gamma}\  \frac{\partial G}{\partial \theta_2} = 0.
\end{eqnarray}
The expressions for the coefficients $A_x^{\gamma}$ and $B^{\gamma}$ are cumbersome and for completeness we collect them in Appendix \ref{1-defo-coef}.
The Virasoro constraints read
\begin{eqnarray}
\kappa^2 &=& 4 G\sin^2\xi\cos^2\xi(\tilde\omega^2+\tilde n^2)
 + G\cos^2\xi\sin^2\theta_1(\omega_2^2+n_2^2)
\nonumber\\
&& 
G\sin^2\xi\sin^2\theta_2(\omega_3^2+n_3^2)+4 \gamma^2G\sin^4\xi\cos^4\xi\sin^2\theta_1\sin^2\theta_2 (\omega_1^2+n_1^2),
\label{vir1defo}
\end{eqnarray}
and
\begin{eqnarray}
\label{vir2defo}
0 &=& 
\sin^2\xi \cos^2\xi \tilde{\omega} \tilde{n}
+\frac{1}{4} \cos^2\xi \sin^2\theta_1 \omega_2 n_2
+\frac{1}{4} \sin^2\xi \sin^2\theta_2 \omega_3 n_3
\nonumber\\
&&
+\gamma^2 \sin^4\xi \cos^4\xi \sin^2\theta_1 \sin^2\theta_2 \omega_1 n_1. 
\end{eqnarray}
The conserved charges relative to this Ansatz are
\begin{eqnarray}
E &=&
\frac{\sqrt{\tilde\lambda}}4\kappa,
\\
J_{\psi}
&=&
\sqrt{\tilde\lambda} G \bigg(\tilde\omega \sin^2\xi\cos^2\xi
+\gamma^2 \omega_1 \sin^4\xi\cos^4\xi\sin^2\theta_1 \sin^2\theta_2
\nonumber\\
&& 
+  \frac{\gamma}{2} \cos^2 \xi\sin^2 \xi \left(n_2 \cos^2\xi\sin^2\theta_1 \cos\theta_2 +n_3 \sin^2\xi\sin^2\theta_2 \cos\theta_1\right)\bigg),
\nonumber \\
\\
J_{\varphi_1}
&=&
\sqrt{\tilde\lambda}\bigg(\frac{1}{2} G \tilde\omega \sin^2\xi\cos^2\xi\cos\theta_1+\frac{\omega_2} 4 G \cos^2\xi\sin^2\theta_1
\nonumber\\
&& 
\quad 
+ \frac{\gamma}{4} \cos^2 \xi\sin^2 \xi G \left(f n_3- 2 n_1 \cos^2\xi\sin^2\theta_1 \cos\theta_2\right)\bigg),
\\
J_{\varphi_2}
&=&
\sqrt{\lambda}\bigg(-\frac{1}{2} G \tilde\omega \sin^2\xi\cos^2\xi\cos\theta_2+\frac{\omega_3} 4 G \sin^2\xi\sin^2\theta_2
\nonumber\\
&&
\quad 
- \frac{\gamma}{4} G  \cos^2 \xi\sin^2 \xi\left(f n_2 + 2 n_1\sin^2\xi\sin^2\theta_2 \cos\theta_1\right)\bigg).
\end{eqnarray}

\subsection{Special solutions}
\label{soluz-speciali}
In this section we are going to consider two possible solutions. 
The first one has no direct analogue in the undeformed case.
The second one is similar to the solution (\ref{case2}) of Section \ref{special-undefo}.

The choices (\ref{soluz1.1}), (\ref{soluz1.2}) still solve the deformed equations of motion and the Virasoro constraints and 
lead exactly to the same conclusions of Section \ref{special-undefo}. 
This means in particular that string states dual to chiral primaries are insensitive to marginal deformations.
This was already observed in \cite{TsT1} where marginal deformations of the $\mathcal{N}=4$ SYM theory where studied.

\begin{itemize}
\item Solution 1 \\
A direct computation can confirm that the following two choices for embedding coordinates, frequencies and mode numbers
\begin{eqnarray}
1.1) &&
\  \, \, \ \  \xi= \frac{\pi}{4}, \quad   \cos \theta_1=-\frac{2 n_1}{n_2}, \quad \theta_2=0,
\nonumber \\
&&
\omega_2=0, \quad \omega_3=0, 
\quad
n_2 = \frac{\gamma\, \omega_1}{2}, \quad  n_3=0,
\nonumber \\
1.2) &&
\  \, \, \ \  \xi= \frac{\pi}{4}, \quad   \cos \theta_1=-\frac{2 n_1}{n_2}, \quad \theta_2=\pi,
\nonumber \\
&&
\omega_2=0, \quad \omega_3=0, 
\quad
n_2 = - \frac{\gamma\, \omega_1}{2}, \quad  n_3=0,
\end{eqnarray}
satisfy the equations of motion (\ref{eomxi}) and the second Virasoro constraint (\ref{vir2defo}).

In both cases, the first Virasoro constraint (\ref{vir1defo}) gives simply
\begin{eqnarray}
 \kappa = \pm \omega_1.
\end{eqnarray}
The conserved charges are
\begin{equation}
\label{cariche-defo}
 E = \frac{\sqrt{\tilde{\lambda}}}{4}\, \kappa,
 \qquad
 J_{\psi} = \mp 2 J_{\varphi_2} = \frac{\sqrt{\tilde{\lambda}}}{4}\,  \omega_1,
 \qquad
 J_{\varphi_1} = \mp \frac{\sqrt{\tilde\lambda}}{4} \, \frac{2 n_1}{\gamma} = \mp \frac{\sqrt{\tilde\lambda}}{8} \, \frac{2 n_1}{n_2} \omega_1.
\end{equation}
Note that if we take the limit $n_1, n_2 \rightarrow 0$ keeping fixed $2 n_1/n_2 \sim 1$, the string becomes the point-like solution (\ref{soluz1.1}), (\ref{soluz1.2})\footnote{
These results are in disagreement with the results for the analogue solutions in the current version of \cite{Wu} (v3).
We are in contact with the author of that paper: A revisited version of \cite{Wu} is in progress where full agreement with our results (\ref{cariche-defo}) is found.
}.

\item Solution 2

Let's consider the solution (\ref{case2}) in this deformed scenario.
So we fix
\begin{equation}
\label{gino}
 \xi=\pi/4, \quad \theta_2=\pi-\theta_1, \quad \omega_2=\omega_3, \quad n_2=n_3.
\end{equation}
We want solutions with $\sin{\theta_1} \neq 0$.
From the equations of motion for $\xi$, $\theta_1$ and $\theta_2$, we get
\begin{equation}
\label{facile}
 \omega_1n_2 = \omega_2n_1,
\end{equation}
and
\begin{eqnarray}
 && 
 \Big(
 \gamma^2 \left(
  \omega_1 \left(\gamma^2 \omega_1 \sin (6\theta_1)-4 \left(\gamma^2+8\right) \omega_1 \sin (4 \theta_1)
  -32 \omega_2 \sin (3 \theta_1)\right)
\right.
\nonumber \\
&&
\qquad \left.
  + \sin (2 \theta_1) \left(5 \gamma^2 \omega_1^2 -64 \omega_2^2\right)
  \right)
 -32 \left(5 \gamma^2+16\right) \omega_1 \omega_2 \sin (\theta_1) \Big) \left(n_1^2- \omega_1^2\right) =0,
 \label{eomdefo}
\nonumber \\
\end{eqnarray}
while the second Virasoro constraints (\ref{vir2defo}) gives
\begin{equation}
\label{viradefo}
n_1 \left(\gamma^2 \omega_1^2 \sin^4(\theta_1)+4 (\omega_2 \cos(\theta_1)+\omega_1)^2+ 4 \omega_2^2 \sin^2(\theta_1)\right) =0.
\end{equation}

In the last two equations we already used condition (\ref{facile}).
Under the requirement that $\sin(\theta_1)$ is nonzero, we can solve these equations in two different ways:
\begin{itemize}

\item Solution 2.1: $\omega_1=\omega_2=0$.
\\
Since all the frequencies are set to zero, the string does not rotate at all.
Thus, this solution is not interesting for our purposes.

\item Solution 2.2: $n_1=n_2=0$.
\\
This is the solution for a point-like string.
\end{itemize}

For this second case, equation (\ref{eomdefo}) gives
\begin{eqnarray}
0 & = &
\gamma^4 \omega_1^2 \sin (6 \theta_1)
-4 \left(\gamma^2+8\right) \gamma^2 \omega_1^2 \sin (4 \theta_1)
-32 \gamma^2 \omega_1 \omega_2 \sin (3 \theta_1)
\nonumber \\
&&
-\gamma^2 \sin (2 \theta_1) \left(64 \omega_2^2 -5 \gamma^2 \omega_1^2\right)
-32 \left(5 \gamma^2+16\right) \omega_1 \omega_2 \sin (\theta_1).
\end{eqnarray}
This is a constraint that can be solved by fixing, for example, the value of one of the frequencies (let's say $\omega_1$) in terms of other variables.

The first Virasoro constraint (\ref{vir1defo}) gives
\begin{equation}
 \kappa^2= \omega_1^2 +\omega_2^2 + 2 \omega_1 \omega_2 \cos(\theta_1)
 -\frac{G}{4} \gamma^2 \sin^2(\theta_1) \left(\omega_1 \cos(\theta_1)+ \omega_2 \right)^2.
\end{equation}
The conserved charges relative to this solution are
\begin{eqnarray} 
\label{charges-defo}
E &=& \frac{\sqrt{\tilde\lambda}}{4} k ,
\nonumber \\
J_{\psi} &=& \frac{\sqrt{\tilde\lambda}}{4} \ G \ ( \omega_1+\cos\theta_1\omega_2 + \gamma^2 \frac{\omega_1 \sin^4 \theta_1}{4})  ,
\nonumber\\
J_{\varphi_1} = J_{\varphi_2} &=&\ \frac{\sqrt{\tilde\lambda}}{8}\ G\ (\omega_1\cos\theta_1+\omega_2),
\end{eqnarray}
with 
\begin{equation}
 G = \frac{4}{4+ \gamma^2 \sin^2(\theta_1)}.
\end{equation}

\end{itemize}

\section{Multi-spin strings in 3-deformed $AdS_4\times CP^3$}
\label{3deformed}

\subsection{General discussion}
In this Section we consider the most general marginal deformation of ABJM theory. It is characterized by 3 real deformation parameters $\gamma_i$, {\footnotesize $(i=1,2,3)$}.
This deformation breaks all the supersymmetry of the original theory but preserves the conformal symmetry.

The potential can be build up from the undeformed superpotential (\ref{superpotential}) following what has been done for the marginal deformations of the $\mathcal{N}=4$ theory \cite{noi}.
The gravity dual of this deformed theory can be found in \cite{Imeroni:2008cr}.
It can be computed starting from the undeformed background (\ref{undef-metric}) by performing this precise sequence of 3 TsT transformations:
$\left(\varphi_1,\varphi_2\right)^{\textrm{TsT}}_{\gamma_3}$, $\left(\varphi_2,\psi\right)^{\textrm{TsT}}_{\gamma_1}$, $\left(\psi,\varphi_1\right)^{\textrm{TsT}}_{\gamma_2}$.

The metric and the NS-NS $B$-field are
\begin{equation}
 ds^2_{\tilde\gamma}=R^2(\frac{1}{4} ds^2_{AdS_4}+ds^2_{CP^3_{\gamma_i}}),
\end{equation}
\begin{eqnarray}
 ds^2_{CP^3_{\tilde\gamma}}&=&
 d\xi^2+\frac{1}{4} \cos^2\xi(d\theta_1^2+G\sin^2\theta_1d\varphi_1^2)+\frac{1}{4} \sin^2\xi(d\theta_2^2 +G\sin^2\theta_2d\varphi_2^2)
\nonumber\\
&&
+G\cos^2\xi\sin^2\xi(d\psi+\frac{1}{2} \cos\theta_1d\varphi_1 -\frac{1}{2} \cos\theta_2d\varphi_2)^2
\nonumber\\
&&
+ G\sin^4\xi\cos^4\xi
\sin^2\theta_1\sin^2\theta_2 (\tilde\gamma_1 d\varphi_1+\tilde\gamma_2 d\varphi_2 + \tilde\gamma_3 d\psi^2)^2 ,
\end{eqnarray}
\begin{eqnarray}
 B &=& -\frac{R^2}{2} \cos^2 \xi\sin^2 \xi G \Big( 
\frac{1}{2} \left(\tilde\gamma_3 \cos\theta_2 +2 \tilde\gamma_2 \right) \cos^2\xi\sin^2\theta_1 d \psi \wedge d \varphi_1 
\nonumber \\
&&
+\frac{1}{2} \left(\tilde\gamma_3 \cos\theta_1 - 2 \tilde\gamma_1 \right) \cos^2\xi\sin^2\theta_2 d \psi \wedge d \varphi_2 
\nonumber \\
&&
+\frac{1}{4} \left(\tilde\gamma_3 \sin^2\theta_1 \sin^2\theta_2 +\left(\tilde\gamma_3 \cos\theta_2+2 \tilde\gamma_2\right)\cos^2\xi\sin^2\theta_1 \cos\theta_2
\right.
\nonumber \\
&&
\qquad
\left.
+\left(\tilde\gamma_3 \cos\theta_1 - 2 \tilde\gamma_1\right)\sin^2\xi\sin^2\theta_2 \cos\theta_1 d\varphi_1 \wedge d\varphi_2\right)
\Big),
\end{eqnarray}
where 
\begin{eqnarray}
\tilde\gamma_i &=& \frac{R^2}{4}\gamma_i,
\nonumber \\
G^{-1} &=& 1 + \cos{\xi}^2 \sin{\xi}^2 \Big(
\tilde{\gamma}_3^2 \sin{\theta_1}^2 \sin{\theta_2}^2
+(\tilde{\gamma}_3 \cos{\theta_2} + 2 \tilde{\gamma}_2)^2 \cos{\xi}^2 \sin{\theta_1}^2 
\nonumber \\
&&
\quad \quad \qquad \qquad \qquad 
+(\tilde{\gamma}_3 \cos{\theta_1} - 2 \tilde{\gamma}_1)^2 \sin{\xi}^2 \sin{\theta_2}^2 
\Big).
\end{eqnarray}
As we did in the previous Section, in the following all the $\gamma_i$ must be considered as $\tilde\gamma_i$ unless explicitly specified.

The Polyakov action in this background is
\begin{eqnarray}
\label{action-defo}
 S &=& \frac{R^2}{16\pi\alpha^\prime}\int d\sigma d\tau
   \Big\{ 
   \nonumber \\
   &&
   -\cosh^2(\rho) \left(t^{\prime 2}-\dot{t}^2\right)
     +\rho^{\prime 2} -\dot{\rho}^2
   +\sinh^2 (\rho) \left(\theta^{\prime 2} -\dot{\theta}^2\right)
   -4 \dot{\xi}^2 +4 \xi^{\prime 2}    
   \nonumber \\
   &&
   +\sin^2(\theta) \sinh^2(\rho) \left(\varphi^{\prime 2} -\dot{\varphi}^2\right)
   +\cos^2(\xi) \left(\theta_1^{\prime 2}- \dot{\theta}_1^2\right)
   +\sin^2(\xi) \left(\theta_2^{\prime 2}   -\dot{\theta}_2^2\right) 
   \nonumber \\
   &&
   + \frac{G}{4} \sin^2(2 \xi) \left(\gamma_3^2 \sin^2(\theta_1) \sin^2(\theta_2) \sin^2(2 \xi)+4\right) \left(\psi^{\prime 2}-\dot{\psi}^2\right) 
   \nonumber \\
   &&
   + G \cos^2(\xi) \left(\sin^2(\theta_1) \left(4 \gamma_1^2 \sin^2(\theta_2) \sin^4(\xi) \cos^2(\xi)+1\right)
   +\cos^2(\theta_1) \sin^2(\xi)\right) \left(\varphi_1^{\prime 2} - \dot{\varphi}_1^2\right) 
   \nonumber \\
   &&
   + G \sin^2(\xi) \left(\sin^2(\theta_2) \left(4 \gamma_2^2 \sin^2(\theta_1) \sin^2(\xi) \cos^4(\xi)+1\right)
   +\cos^2(\theta_2) \cos^2(\xi)\right) \left(\varphi_2^{\prime 2}-\dot{\varphi}_2^2\right) 
   \nonumber \\
   &&
   + G \sin^2(2 \xi) \left(\frac{1}{2} \gamma_1 \gamma_3 \sin^2(\theta_1) \sin^2(\theta_2) \sin^2(2 \xi)+\cos(\theta_1)\right)
   \left(\psi^{\prime} \varphi_1^{\prime}-\dot{\psi} \dot{\varphi}_1\right) 
   \nonumber \\
   &&
   - G \sin^2(2 \xi) \left(\cos (\theta_2)-\frac{1}{2} \gamma_2 \gamma_3 \sin^2(\theta_1) \sin^2(\theta_2) \sin^2(2 \xi)\right)
   \left(\psi^{\prime} \varphi_2^{\prime} -\dot{\psi} \dot{\varphi}_2\right) 
   \nonumber \\
   &&
   + \frac{G}{2} \sin^2(2 \xi) \left(\gamma_1 \gamma_2 \sin^2(\theta_1) \sin^2(\theta_2) \sin^2(2 \xi)
   -\cos (\theta_1) \cos (\theta_2)\right) \left(\varphi_1^{\prime} \varphi_2^{\prime}-\dot{\varphi}_1 \dot{\varphi}_2\right) 
   \nonumber \\
   &&
   + G \sin^2(\theta_1) \sin^2(2 \xi) \cos^2(\xi) (2 \gamma_2+\gamma_3 \cos(\theta_2)) \left(\psi^{\prime} \dot{\varphi}_1-\dot{\psi} \varphi_1^{\prime}\right)
   \nonumber \\
   &&
   + G \sin^2(\theta_2) \sin^2(\xi) \sin^2(2 \xi) (\gamma_3 \cos(\theta_1) -2 \gamma_1) \left(\psi^{\prime} \dot{\varphi}_2 - \dot{\psi} \varphi_2^{\prime}\right)
   \nonumber \\
   &&
   + \frac{G}{2} \sin^2(2 \xi) \left(\varphi_1^{\prime} \dot{\varphi}_2-\dot{\varphi}_1 \varphi_2^{\prime}\right) 
   \left(\cos(\theta_1) \sin^2(\theta_2) \sin^2(\xi) (\gamma_3 \cos(\theta_1) - 2 \gamma_1)
   \right.
   \nonumber \\
   &&
   \left.
   \qquad 
   +\sin^2(\theta_1) \cos(\theta_2) \cos^2(\xi) (2 \gamma_2+ \gamma_3 \cos(\theta_2))+\gamma_3 \sin^2(\theta_1) \sin^2(\theta_2)\right)
\Big\}.
\end{eqnarray}
A direct check shows that this expression reduces to the $\beta$-deformed action (\ref{azione-1defo}) if we fix $\gamma_1=\gamma_2=0$ and $\gamma_3=\gamma$.

By using the Ansatz (\ref{ansatz}) and requiring $\xi$, $\theta_1$ and $\theta_2$ to be constants,
it is immediately seen that the equations of motion relative to the coordinates $t$, $\rho$, $\theta$, $\varphi$, $\psi$, $\varphi_1$ and $\varphi_2$ are trivially satisfied.

The equations of motion for $\xi$, $\theta_1$ and $\theta_2$ read
\begin{eqnarray}
\label{eomxi-defo}
&&
\qquad \qquad \frac{1}{2}\ \cos\xi \sin\xi\  A_{\xi}^{\gamma_i}\, G + \frac{B^{\gamma_i}}{256}\  \frac{\partial G}{\partial \xi} =0,
\\
\nonumber \\
\label{eomth1-defo}
&& \frac{A^{\gamma_i}_{\theta_1}}{16} G + \frac{B^{\gamma_i}}{256} \frac{\partial G}{\partial \theta_1} = 0,
\qquad \qquad 
\frac{A^{\gamma_i}_{\theta_2}}{16} G + \frac{B^{\gamma_i}}{256} \frac{\partial G}{\partial \theta_2} = 0.
\label{eomth2-defo}
\end{eqnarray}
The coefficients $A_x^{\gamma_i}$ and $B^{\gamma_i}$ can be found in Appendix \ref{3-defo-coef}.
The Virasoro constraints read
\begin{eqnarray}
\label{Vira1-defo}
\kappa ^2 &=&
   G \sin^2(\theta_1) \cos^2(\xi) \left(n_2^2+\omega_2^2\right)
+ G \sin^2(\theta_2) \sin^2(\xi) \left(n_3^2+\omega_3^2\right)
+ G \sin^2(2 \xi) \left(\tilde{n}^2+\tilde{\omega}^2\right)
\nonumber \\
&&
+\frac{G}{4} \sin ^2(\theta_1) \sin ^2(\theta_2) \sin^4(2 \xi) 
\times 
\nonumber \\
&&
\qquad
\times
\left(\gamma_3^2 \left(n_1^2+\omega_1^2\right)+\gamma_1^2 \left(n_2^2+\omega_2^2\right)+\gamma_2^2
   \left(n_3^2+\omega_3^2\right)
   \right.
   \nonumber \\
   &&
   \qquad \quad
   \left.
   + 2 \gamma_1 \gamma_3 (n_1 n_2+\omega_1 \omega_2)
   +2 \gamma_2 \gamma_3 (n_1 n_3+\omega_1\omega_3)
   +2 \gamma_1 \gamma_2 (n_2 n_3+\omega_2 \omega_3)
  \right),
\end{eqnarray}
and
\begin{eqnarray}
\label{Vira2-defo}
0 &=& 
\left(n_2 \omega_2 \sin^2(\theta_1) \cos^2(\xi)+n_3 \omega_3 \sin^2(\theta_2)
   \sin^2(\xi)+\tilde{n} \tilde{\omega} \sin^2(2 \xi)\right)
\nonumber \\
&&
+ 4 \sin^2(\theta_1) \sin^2(\theta_2) \sin^4(\xi) \cos^4(\xi) (\gamma_1 \omega_2+\gamma_2 \omega_3+\gamma_3 \omega_1) (\gamma_3 n_1+\gamma_1 n_2+\gamma_2 n_3).
\nonumber \\
\end{eqnarray}
The charges are
\begin{eqnarray}
E &=&
\frac{\sqrt{\tilde\lambda}}4\kappa,
\end{eqnarray}
\begin{eqnarray}
J_{\psi}
&=&
\sqrt{\tilde\lambda} G \bigg( \tilde\omega \sin^2\xi\cos^2\xi +\frac{1}{2} \sin^2(\xi) \cos^2(\xi) 
\times 
\nonumber\\
&&
\quad 
\times
\left(\sin^2(\theta_1) \cos^2(\xi) \left(2 \gamma_3 \sin^2(\theta_2) \sin^2(\xi) (\gamma_1 \omega_2+\gamma_2 \omega_3+\gamma_3 \omega_1)
+2 \gamma_2 n_2+\gamma_3 n_2 \cos(\theta_2)\right)
\right.
\nonumber \\
&&
\left.
\qquad
+n_3 \sin^2(\theta_2) \sin^2(\xi) (\gamma_3 \cos(\theta_1)-2 \gamma_1)\right)
\bigg),
\end{eqnarray}
\begin{eqnarray}
J_{\varphi_1}
&=&
\sqrt{\tilde\lambda} G \bigg(\frac{1}{2} \tilde\omega \sin^2\xi\cos^2\xi\cos\theta_1+\frac{\omega_2} 4 \cos^2\xi\sin^2\theta_1
\nonumber\\
&& 
\qquad
-\frac{1}{2} \sin^2(\theta_1) \sin ^2(\xi ) \cos ^4(\xi ) (2 \gamma_2 n_1+\cos (\theta_2) (\gamma_3 n_1-\gamma_2 n_3))
   \nonumber \\
   &&
   \qquad
   +\frac{1}{4} \sin ^2(\theta_2) \sin ^4(\xi ) \cos ^2(\xi ) \left(4 \gamma_1 \sin ^2(\theta_1) \cos ^2(\xi ) (\gamma_1 \omega_2+\gamma_2 \omega_3+\gamma_3 \omega_1)
   \right.
   \nonumber \\
   &&
   \left.
   \qquad \qquad 
   +n_3 \cos(\theta_1) (\gamma_3 \cos (\theta_1)-2 \gamma_1)\right)   
   \nonumber \\
   &&
   \qquad
   +\frac{1}{4} \gamma_3 n_3 \sin ^2(\theta_1) \sin ^2(\xi ) \cos ^2(\xi ) \left(\cos ^2(\theta_2) \cos ^2(\xi )+\sin ^2(\theta_2)\right)
\bigg),
\end{eqnarray}
\begin{eqnarray}
J_{\varphi_2}
&=&
\sqrt{\tilde\lambda} G \bigg(-\frac{1}{2} \tilde\omega \sin^2\xi\cos^2\xi\cos\theta_2+\frac{\omega_3} 4 \sin^2\xi\sin^2\theta_2
\nonumber\\
&&
\qquad
+\gamma_1 n_1 \sin ^2(\theta_2) \sin ^4(\xi ) \cos ^2(\xi)
\nonumber \\
&&
\qquad
+\frac{1}{4} \sin ^2(\theta_2) \sin ^4(\xi ) \cos ^2(\xi ) \left(4 \gamma_2 \sin ^2(\theta_1) \cos ^2(\xi ) (\gamma_1 \omega_2+\gamma_2 \omega_3+\gamma_3 \omega_1)
\right.
\nonumber \\
&&
\left.
\qquad \qquad
-\cos (\theta_1) (2 \gamma_3 n_1-2 \gamma_1 n_2+\gamma_3 n_2 \cos (\theta_1))\right)
\nonumber \\
&&
\qquad
-\frac{1}{16} n_2 \sin ^2(\theta_1) \sin ^2(2 \xi ) \left(\cos (\theta_2) \cos ^2(\xi ) (2 \gamma_2+\gamma_3 \cos (\theta_2))+\gamma_3 \sin ^2(\theta_2)\right)
\bigg)
.
\nonumber \\
\end{eqnarray}

\subsection{Special solutions}
Let's consider in this scenario the two solutions presented in Section \ref{soluz-speciali} for the one-deformation parameter case.
The results we find here are a generalization to the non-supersymmetric case of the $\beta$-deformed solutions.

\begin{itemize}
\item Solution 1 \\
One can confirm by direct inspection that the following two choices for embedding coordinates, frequencies and mode numbers
\begin{eqnarray}
1.1) &&
\  \, \, \ \ \ \ \ \  \xi= \frac{\pi}{4}, \quad   \cos \theta_1=-\frac{2 n_1}{n_2}, \quad \theta_2=0,
\nonumber \\
&&
\omega_2=0, \quad \omega_3=0, 
\quad
n_2 = \frac{\omega_1}{2} (2 \gamma_2 + \gamma_3), \quad  n_3=0,
\nonumber \\
1.2) &&
\  \, \, \ \ \ \ \ \  \xi= \frac{\pi}{4}, \quad   \cos \theta_1=-\frac{2 n_1}{n_2}, \quad \theta_2=\pi,
\nonumber \\
&&
\omega_2=0, \quad \omega_3=0, 
\quad
n_2 = - \frac{\omega_1}{2} (2 \gamma_2 - \gamma_3) , \quad  n_3=0,
\end{eqnarray}
satisfy the equations of motion (\ref{eomxi-defo}), (\ref{eomth1-defo}) and the second Virasoro constraint (\ref{Vira2-defo}).

For both the solutions, the first Virasoro constraint (\ref{Vira1-defo}) gives simply
\begin{eqnarray}
 \kappa = \pm \omega_1.
\end{eqnarray}
The conserved charges are
\begin{equation}
 E = \frac{\sqrt{\tilde{\lambda}}}{4}\, \kappa,
 \quad
 J_{\psi} = \mp 2 J_{\varphi_2} = \frac{\sqrt{\tilde{\lambda}}}{4}\,  \omega_1,
 \quad
 J_{\varphi_1} = \frac{\sqrt{\tilde\lambda}}{4} \, \frac{2 n_1}{2 \gamma_2 \pm \gamma_3}=\mp \frac{\sqrt{\tilde\lambda}}{8} \, \frac{2 n_1}{n_2} \omega_1.
\label{risultato}
\end{equation}
Correctly, by fixing $\gamma_2=0$, $\gamma_3 = \gamma$, these results reduce to the ones we found in the one-deformation parameter case (see eq. (\ref{cariche-defo})). 
We note that the charges (\ref{risultato}) are independent of $\gamma_1$. Similar behaviors of the deformation parameters have already been observed in the literature on marginal deformations (see, for example, \cite{noi}, \cite{pirrone} where brane solutions in marginal deformed $AdS_5\times S^5$ are considered).

\item Solution 2

Let's reconsider the Ansatz (\ref{gino}) in the three-deformation parameter background. 
The Ansatz was
\begin{equation}
\label{gino2}
 \xi=\pi/4, \quad \theta_2=\pi-\theta_1, \quad \omega_2=\omega_3, \quad n_2=n_3.
\end{equation}
We remind that we are interested only in solutions with $\sin{\theta_1} \neq 0$.

In order to simplify the equations of motion and the Virasoro constraints as much as possible, we require also 
\begin{equation}
\label{mah}
n_2 = n_1 =0.
\end{equation}
This condition was also required in the $\beta$-deformed scenario.

Despite this simplification, the equations are still extremely complicated.
Stronger simplifications which lead us to our final results occur if we fix $\gamma_1 = \gamma_2$.
In particular, the equations of motion for the $\xi$ variable (\ref{eomxi-defo}) and the second Virasoro constraint (\ref{Vira2-defo}) are directly satisfied.
We remark that this choice does not restore any supersymmetry.

The equations of motion associated to $\theta_1$ and $\theta_2$ give the same constraint on the free parameters, namely
\begin{eqnarray}
   0 &=& 
   8 \gamma_2^2 \omega_1^2 \sin (2 \theta_1)+\gamma_3^2 \omega_2^2 \sin ^3(2 \theta_1)
   + 3 \gamma_3^2 \omega_1 \omega_2 \sin (\theta_1) \sin ^2(2 \theta_1)
   \nonumber \\
   &&   
   + \gamma_3 \sin ^7(\theta_1) (2 \gamma_2-\gamma_3 \cos (\theta_1)) (2 \gamma_2 \omega_2+\gamma_3 \omega_1)^2 
   \nonumber \\
   &&
   +\sin ^5(\theta_1) \Big(4 \gamma_3 \omega_2 (2 \gamma_2 \omega_2+\gamma_3 \omega_1)
   \nonumber \\
   &&
   \quad \quad
   -\cos (\theta_1)
   \left(\left(4 \gamma_2^2 \omega_2-\gamma_3 \cos (\theta_1) (2 \gamma_2 \omega_2+\gamma_3 \omega_1)+2
   \gamma_2 \gamma_3 \omega_1\right)^2-4 \gamma_3^2 \omega_2^2\right)\Big)
   \nonumber \\
   &&
   + 4 \sin ^3(\theta_1) \Big(2 \gamma_2 \omega_1 (2 \gamma_2 \omega_2+\gamma_3 \omega_1)
   \nonumber \\
   &&
   \quad \quad
   -\cos (\theta_1)
   \left(4 \gamma_2^2 \omega_2^2+2 \gamma_2 \gamma_3 \omega_2^2 \cos (\theta_1)+8 \gamma_2 \gamma_3
   \omega_1 \omega_2+\gamma_3^2 \omega_1^2\right)\Big)
   \nonumber \\
   &&
   +16 \omega_1 \omega_2 \sin (\theta_1)
   +4 \sin (\theta_1) \cos ^3(\theta_1) (-2 \gamma_2 \omega_2+\gamma_3 \omega_2 \cos (\theta_1)+\gamma_3
   \omega_1)^2  
   \nonumber \\
   &&
   -16 \gamma_2 \omega_1 \sin (\theta_1) \cos ^2(\theta_1) (-2 \gamma_2 \omega_2+\gamma_3 \omega_2 \cos
   (\theta_1)+\gamma_3 \omega_1).
\end{eqnarray}
This equation is solved by suitably fixing $\omega_1$ (or $\omega_2$) as a function of the other parameters.

The first Virasoro constraint is 
\begin{eqnarray}
   k^2 &=& \frac{N}{D},
   \\
   N &=& 
   -4 \left(3 \gamma_2^2+8\right) \omega_2^2
   -(\cos (4 \theta_1)-4 \cos (2 \theta_1)) (2 \gamma_2 \omega_2+\gamma_3\omega_1)^2
   \nonumber \\
   &&
   \qquad
   -12 \gamma_2 \gamma_3 \omega_1 \omega_2
   -\left(3 \gamma_3^2+32\right) \omega_1^2
   -64 \omega_1\omega_2 \cos (\theta_1),
   \nonumber \\
   D &=& 
   4 \left(4 \gamma_2^2+\gamma_3^2\right) \cos (2 \theta_1)
   -4 \left(4 \gamma_2^2+2 \gamma_2 \gamma_3 \cos (3 \theta_1)+\gamma_3^2+8\right)
   +8 \gamma_2 \gamma_3 \cos (\theta_1).
   \nonumber
\end{eqnarray}
This condition fixes the value of $k$ to the value of the other parameters.

The values for the charges associated to this solution are
\begin{eqnarray}
E &=&
\frac{\sqrt{\tilde\lambda}}{4}\ \kappa,
\\
J_{\psi}
&=&
\sqrt{\tilde\lambda} \ 
\frac{\gamma_3 \sin ^4(\theta_1) (2 \gamma_2 \omega_2+\gamma_3 \omega_1)+4 \omega_2 \cos (\theta_1)+4
   \omega_1}{4 \sin ^2(\theta_1) \left(4 \gamma_2^2-4 \gamma_2 \gamma_3 \cos (\theta_1)+\gamma_3^2\right)+16},
\\
J_{\varphi_1} = J_{\varphi_2}
&=&
\sqrt{\tilde\lambda}\ 
\frac{\gamma_2 \sin ^4(\theta_1) (2 \gamma_2 \omega_2+\gamma_3 \omega_1)+2 \omega_1 \cos (\theta_1) +2 \omega_2 }
{4 \sin ^2(\theta_1) \left(4 \gamma_2^2-4 \gamma_2 \gamma_3 \cos (\theta_1)+\gamma_3^2\right)+16}.
\end{eqnarray}
By fixing $\gamma_1 = \gamma_2 =0$ and $\gamma_3 = \gamma$ these charges match precisely the value of the charges (\ref{charges-defo}) we found in the $\beta$-deformed case.
Fixing also $\gamma_3=0$ we recover the undeformed results (\ref{cariche-undefo})\footnote{Of course, this happens after we set to zero the mode numbers in (\ref{cariche-undefo}) consistently with (\ref{mah}).}.

\end{itemize}

\section{Conclusion and discussion}

In this paper, we focused on multi-spin string solutions in $AdS_4\times CP^3$ and its marginal deformations.
We considered both a supersymmetric one-deformation parameter background and a non-supersymmetric 3-deformation parameter one.
The first case was already considered in \cite{Wu}.
Our solutions reproduces those results and generalizes them to the 3-deformation parameter scenario. 

In view of further checks of the $AdS/CFT$ correspondence in cases with reduced supersymmetry, it could be interesting to find other string solutions in marginally deformed backgrounds.
Of greatest interest for their relations with the Konishi-like operators could be the study of the folded string in these marginally deformed backgrounds.
This could be interesting also for a comparison of that solution with analogue folded string solutions already found in marginally deformed $AdS_5 \times S^5$ \cite{folded-marg}.
It would also be interesting to generalize the giant magnon and spiky string solutions in $AdS_4 \times CP^3$ discussed in \cite{ref-add:Giardino} and \cite{giant} to the non-supersymmetic marginal deformed case.

This paper offers the starting point to study all these solutions, i.e. the general expressions for the actions (\ref{azione-1defo}) and (\ref{action-defo}). 
We hope to come back to these issues in the near future.

\vskip 25pt
\section*{Acknowledgements}
\noindent
This work has been supported in part by INFN and MIUR.
\\
I would like to thank Marc Grisaru for useful discussion and continuous encouragements
and Silvia Penati, Gabriele Tartaglino Mazzucchelli and Sean T. McReynolds for carefully reading the paper and giving helpful suggestions.

\vfill
\newpage
\appendix
\section{Details on the coefficients in the equations of motion}
\subsection{One-deformation parameter coefficients}
\label{1-defo-coef}
\begin{eqnarray}
 A_{\xi}^{\gamma} &=& 
 2 \cos^2(\theta_1 ) \sin (4 \xi) (n_2-\omega_2) (n_2+\omega_2)
 \nonumber \\
 &&
 +2 \cos^2(\theta_2) \sin (4 \xi) (n_3-\omega_3) (n_3+\omega_3)
 \nonumber \\
 &&
 +8 \sin (\xi) \cos (\xi) \left(\sin^2(\theta_1 ) \left(\omega_2^2-n_2^2\right)+\sin^2(\theta_2) (n_3-\omega_3)
   (n_3+\omega_3)\right)
 \nonumber \\
 &&
 +8 \cos (\theta_2) \sin (2 \xi) \cos (2 \xi) \left(\cos (\theta_1 ) (n_2 n_3-\omega_2 \omega_3)+\cos (\theta_2)
   \left(\omega_3^2-n_3^2\right)-2 n_3 \tilde{n}+2 \omega_3 \tilde{\omega}\right)
 \nonumber \\
 &&
 -4 \cos^2(\theta_1 ) \sin (4 \xi) (n_2 (n_2+\tilde{n})-\omega_2 (\omega_2+\tilde{\omega}))
 \nonumber \\
 &&
 + 2 \sin (4 \xi) (\cos (\theta_1 ) (\omega_2-n_2)+\cos (\theta_2) (n_3-\omega_3)+2 (\tilde{n}-\tilde{\omega})) 
 \times
 \nonumber \\
 && 
\qquad \qquad \times 
 (\cos (\theta_2) (n_3+\omega_3)-\cos(\theta_1 ) (n_2+\omega_2)+2 (\tilde{n}+\tilde{\omega}))
 \nonumber \\
 &&
 +\gamma  \cos^2(\theta_1 ) \sin (4 \xi) (n_3 (\omega_2+\tilde{\omega})-\omega_3 (n_2+\tilde{n}))
 \nonumber \\
 &&
 -2 \gamma  (n_3 \omega_2-n_2 \omega_3) \left(\sin^2(2 \xi) f^{(1,0,0)}(\xi,\theta_1 ,\theta_2)+2 \sin (4 \xi) f(\xi
   ,\theta_1 ,\theta_2)\right)
 \nonumber \\
 &&
 -\gamma  \cos (\theta_1 ) \sin (\xi) \left(6 \cos (2 \theta_2) \sin^2(\xi) \cos (3 \xi)+\cos (\xi) \left(10 \cos (2 \theta_2) \sin^2(\xi)+3 \cos (4
   \xi)+1\right)\right) \times
   \nonumber \\
   &&
   \qquad \qquad 
\times    
    (\cos (\theta_1 ) (n_3 \omega_2-n_2 \omega_3)+2 (\tilde{n} \omega_3- n_3 \tilde{\omega}))
 \nonumber \\
 &&
 + 2 \gamma  \sin^2(\theta_1 ) \cos (\theta_2) \sin (2 \xi) \cos^2(\xi) (3 \cos (2 \xi)-1) (\cos (\theta_2) (n_3 \omega_2-n_2
   \omega_3)+2 (\tilde{n} \omega_2-n_2 \tilde{\omega}))
 \nonumber \\
 &&
 +2 \gamma^2 \sin^2(\theta_1 ) \sin^2(\theta_2) \sin^3(2 \xi) \cos (2 \xi) (\cos (\theta_1 ) (\omega_2-n_2)+\cos (\theta_2)
   (n_3-\omega_3)+2 (\tilde{n}-\tilde{\omega}))
   \times
\nonumber \\
&&
\qquad \qquad 
\times    
    (-\cos (\theta_1 ) (n_2+\omega_2)+\cos (\theta_2) (n_3+\omega_3)+2 (\tilde{n}+\tilde{\omega})).
\end{eqnarray}
\begin{eqnarray}
A_{\theta_1}^{\gamma} &=&
2\sin (2 \theta_1) \cos^2(\xi) (n_2^2-\omega_2^2)
\nonumber \\
&&
+ 2\sin (\theta_1) \sin^2(2 \xi) (\omega_2 \tilde{\omega}-n_2 \tilde{n})
\nonumber \\
&&
+\gamma  \sin^2(2 \xi) (n_2 \omega_3-n_3 \omega_2) \frac{\partial f}{\partial \theta_1}
\nonumber \\
&&
+ 4 \gamma  \sin (\theta_1) \sin^2(\theta_2) \sin^4(\xi) \cos^2(\xi) (\cos (\theta_1) (n_2 \omega_3-n_3 \omega_2)
+2 (n_3 \tilde{\omega}-\tilde{n} \omega_3))
\nonumber \\
&&
+ 4 \gamma  \sin (2 \theta_1) \cos (\theta_2) \sin^2(\xi) \cos^4(\xi) (\cos (\theta_2) (n_3 \omega_2-n_2 \omega_3)
+2 (\tilde{n} \omega_2-n_2 \tilde{\omega}))
\nonumber \\
&& 
+ 2 
\gamma^2 \sin (2 \theta_1) \sin^2(\theta_2) \sin^4(\xi) \cos^4(\xi) \times 
\nonumber \\
&&
\qquad \quad
\times
(\cos (\theta_1) (\omega_2-n_2)+\cos (\theta_2)
   (n_3-\omega_3)+2 (\tilde{n}-\tilde{\omega}))
   \times
   \nonumber \\
   &&
   \qquad \quad
   \times 
    (\cos (\theta_2) (n_3+\omega_3)-\cos (\theta_1) (n_2+\omega_2)+2 (\tilde{n}+\tilde{\omega})).
\end{eqnarray}
\begin{eqnarray}
A_{\theta_2}^{\gamma} &=&
2 \sin (2 \theta_2) \sin^2(\xi) \left(n_3^2-\omega_3^2\right)
\nonumber \\
&&
+ 2 \sin (\theta_2) \sin^2(2 \xi) (n_3 \tilde{n}-\omega_3 \tilde{\omega})
\nonumber \\
&&
+ \gamma  \sin^2(2 \xi) (n_2 \omega_3-n_3 \omega_2) \frac{\partial f}{\partial \theta_2}
\nonumber \\
&&
+ 4 \gamma  \sin^2(\theta_1) \sin (\theta_2) \sin^2(\xi) \cos^4(\xi) (\cos (\theta_2) (n_2 \omega_3-n_3 \omega_2)+2
   (n_2 \tilde{\omega}-\tilde{n} \omega_2))
\nonumber \\
&&
+ 4 \gamma  \cos (\theta_1) \sin (2 \theta_2) \sin^4(\xi) \cos^2(\xi) (\cos (\theta_1) (n_3 \omega_2-n_2 \omega_3)+2
   (\tilde{n} \omega_3-n_3 \tilde{\omega}))   
\nonumber \\
&&
+ 2 \gamma^2 \sin^2(\theta_1) \sin (2 \theta_2) \sin^4(\xi) \cos^4(\xi) \times
\nonumber \\
&&
\qquad \quad
\times
(\cos (\theta_1) (\omega_2-n_2)+\cos (\theta_2)
   (n_3-\omega_3)+2 (\tilde{n}-\tilde{\omega}))
   \times
   \nonumber \\
   &&
   \qquad \quad
   \times 
    (\cos (\theta_2) (n_3+\omega_3)-\cos (\theta_1) (n_2+\omega_2)+2 (\tilde{n}+\tilde{\omega})).
\end{eqnarray}
\begin{eqnarray}
 B^{\gamma} &=& 
 4 \sin^2(\theta_1) \cos^2(\xi) \left(n_2^2-\omega_2^2\right)+4 \sin^2(\theta_2) \sin^2(\xi) \left(n_3^2-\omega_3^2\right)+4
   \sin^2(2 \xi) \left(\tilde{n}^2-\tilde{\omega}^2\right)
   \nonumber \\
   && 
   + 2 \gamma  \sin^2(2 \xi) (n_2 \omega_3-n_3 \omega_2) f
      \nonumber \\
   && 
   + 8 \gamma  \cos (\theta_1) \sin^2(\theta_2) \sin^4(\xi) \cos^2(\xi) (\cos (\theta_1) (n_3 \omega_2-n_2 \omega_3)
   -2 n_3 \tilde{\omega}+2 \tilde{n} \omega_3)
   \nonumber \\
   &&
   + 8 \gamma  \sin^2(\theta_1) \cos (\theta_2) \sin^2(\xi) \cos^4(\xi) (\cos (\theta_2) (n_3 \omega_2-n_2 \omega_3)
   -2 n_2 \tilde{\omega}+2 \tilde{n} \omega_2)
   \nonumber \\
   &&
   + 4 \gamma^2 \sin^2(\theta_1) \sin^2(\theta_2) \sin^4(\xi) \cos^4(\xi) (\cos (\theta_1) (\omega_2-n_2)+\cos (\theta_2)
   (n_3-\omega_3)+2 (\tilde{n}-\tilde{\omega})) 
   \times
   \nonumber \\
   &&
   \qquad \qquad \times 
   (\cos (\theta_2) (n_3+\omega_3)-\cos (\theta_1) (n_2+\omega_2)+2 (\tilde{n}+\tilde{\omega})).
\end{eqnarray}
\newpage

\subsection{3-deformation parameter coefficients}
\label{3-defo-coef}
\begin{eqnarray}
 A_{\xi}^{\gamma_i} &=& 
 \sin^2(\theta_1) \left(\omega_2^2-n_2^2\right)+\sin^2(\theta_2) \left(n_3^2-\omega_3^2\right)+4 \cos (2 \xi)
   \left(\tilde{n}^2-\tilde{\omega}^2\right)
   \nonumber \\ 
   && 
   +4 \gamma_3 \sin^2(\theta_1) \sin^2(\theta_2) \cos (\theta_2) \sin^2(\xi) \cos^4(\xi) 
   \left(2 \gamma_1 (n_2 n_3-\omega_2 \omega_3)
   \right.
   \nonumber \\
   &&
   \left.
   \qquad \qquad
   +\gamma_3 \cos (\theta_1) (\omega_2 \omega_3-n_2 n_3)
   +2 \gamma_2 \left(n_3^2 -\omega_3^2\right) +2 \gamma_3 (n_3 \tilde{n}-\omega_3 \tilde{\omega})\right)
   \nonumber \\
   &&
   + 4 \gamma_3 \sin^2(\theta_1) \cos(\theta_2) \cos^4(\xi) (\tilde{n} \omega_2-n_2 \tilde{\omega})
   +8 \gamma_2 \sin^2(\theta_1) \cos^4(\xi) (\tilde{n} \omega_2-n_2 \tilde{\omega})
   \nonumber \\
   &&
   +2 \gamma_3^2 \sin^2(\theta_1) \sin^2(\theta_2) \cos^2(\theta_2) \sin^2(\xi) \cos^4(\xi) \left(n_3^2-\omega_3^2\right)
   \nonumber \\
   &&
   + 2 \sin^2(\theta_1) \left(\gamma_3 \sin^2(\theta_2) \sin^2(\xi) (n_3 \omega_2-n_2 \omega_3)
   +2 \gamma_2 \sin^2(2 \xi) (n_2 \tilde{\omega}-\tilde{n} \omega_2)\right)
   \nonumber \\
   &&
   + \sin^2(\theta_2) \sin^2(\xi) (n_3 \tilde{\omega}-\tilde{n} \omega_3) \left(12 \gamma_1 \cos(2 \xi)+4 \gamma_1
   \right.
   \nonumber \\
   &&
   \left.
   \qquad \qquad
   -\gamma_3 \cos (\theta_1-2 \xi)-\gamma_3 \cos (\theta_1+2 \xi)+2 \gamma_3 \cos (\theta_1)\right)
   \nonumber \\
   &&
   -2 \gamma_3 \sin^2(\theta_1) \sin^2(\theta_2) \cos^2(\xi) (n_3 \omega_2-n_2 \omega_3)
   -2 \gamma_3 \sin^2(\theta_1) \cos(\theta_2) \sin^2(2 \xi) (\tilde{n} \omega_2-n_2 \tilde{\omega})
   \nonumber \\
   &&
   -2 \gamma_3 \cos(\theta_1) \sin^2(\theta_2) \sin^2(2 \xi) (n_3 \tilde{\omega}-\tilde{n} \omega_3)
   \nonumber \\
   &&
   +\sin^2(\theta_1) \sin^2(\theta_2) \sin^2(\xi) \cos^4(\xi) 
   \times 
   \nonumber \\
   &&
   \qquad \times
   \left(8 \gamma_1^2 \left(n_2^2-\omega_2^2\right)
   +\gamma_3^2 (2 \cos (2 \theta_1)+1) \left(n_2^2-\omega_2^2\right)
   \right.
   \nonumber \\
   &&
   \left.
   \qquad \quad 
   -8 \gamma_3 \cos (\theta_1) \left(\gamma_1 \left(n_2^2-\omega_2^2\right)
   +\gamma_2 (n_2 n_3-\omega_2 \omega_3)
   +\gamma_3 (n_2 \tilde{n}-\omega_2 \tilde{\omega})\right)
   \right.
   \nonumber \\
   &&
   \left.
   \qquad \quad    
   +16 \gamma_1 \gamma_2 (n_2 n_3-\omega_2 \omega_3)
   +16 \gamma_1 \gamma_3 (n_2\tilde{n}-\omega_2 \tilde{\omega})
   +16 \gamma_2 \gamma_3 (n_3\tilde{n}-\omega_3 \tilde{\omega})
   \right.
   \nonumber \\
   &&
   \left.
   \qquad \quad    
   +8 \gamma_2^2 \left(n_3^2-\omega_3^2\right)
   +8 \gamma_3^2 \left(\tilde{n}^2-\tilde{\omega}^2\right)
   \right)
   \nonumber \\
  &&
  -2 \sin^2(\theta_1) \sin^2(\theta_2) \sin^4(\xi) \cos^2(\xi) 
   \times 
   \nonumber \\
   &&
   \qquad \times
  \left(4 \gamma_1^2 \left(n_2^2-\omega_2^2\right)
    +4 \gamma_2^2 \left(n_3^2-\omega_3^2\right)
  +4 \gamma_3^2 \left(\tilde{n}^2-\tilde{\omega}^2\right)
\right.
\nonumber \\
&&
\left.  
\qquad \quad
  +\gamma_3^2 \cos^2(\theta_1) \left(n_2^2-\omega_2^2\right)
  +\gamma_3^2 \cos^2(\theta_2) \left(n_3^2-\omega_3^2\right)
\right.
\nonumber \\
&&
\left.
\qquad \quad
    -2 \gamma_3^2 \cos (\theta_1) \cos (\theta_2) (n_2 n_3-\omega_2 \omega_3)
   \right.
   \nonumber \\
   &&
   \left.
   \qquad \quad    
  -4 \gamma_3 \cos (\theta_1) \left(\gamma_1 \left(n_2^2-\omega_2^2\right)
  +\gamma_2 (n_2 n_3-\omega_2 \omega_3)
  +\gamma_3 (n_2 \tilde{n}-\omega_2 \tilde{\omega})\right)
   \right.
   \nonumber \\
   &&
   \left.
   \qquad \quad    
  +4 \gamma_3 \cos (\theta_2) \left(\gamma_1 (n_2 n_3-\omega_2 \omega_3)
  +\gamma_2 \left(n_3^2-\omega_3^2\right)+\gamma_3 (n_3 \tilde{n}-\omega_3 \tilde{\omega})\right)
    \right.
   \nonumber \\
   &&
   \left.
   \qquad \quad    
 +8 \gamma_1 \gamma_2 (n_2 n_3-\omega_2 \omega_3) 
  +8 \gamma_1 \gamma_3 (n_2 \tilde{n}-\omega_2 \tilde{\omega})
  +8 \gamma_2 \gamma_3 (n_3\tilde{n}-\omega_3 \tilde{\omega})
  \right).
\end{eqnarray}
\vspace{2mm}
\begin{eqnarray}
\label{eomth1-coef-defo}
A_{\theta_1}^{\gamma_i} &=& 
   8 \sin(\theta_1) \cos^2(\xi) \left(\cos(\theta_1) \left(n_2^2-\omega_2^2\right)+2 \sin^2(\xi) (\omega_2 \tilde{\omega}-n_2 \tilde{n})\right)
   \nonumber \\
   &&
   +2 \gamma_3 \sin (2 \theta_1) \sin^2(\theta_2) \sin^2(2 \xi) (n_2 \omega_3-n_3 \omega_2)
   \nonumber \\
   &&
   +\sin (2 \theta_1) \sin^2(\theta_2) \sin^4(2 \xi) 
   \times
   \nonumber \\
   &&
   \qquad
   \times
    \left(\gamma_1^2 \left(n_2^2-\omega_2^2\right)
   +\gamma_2^2 \left(n_3^2-\omega_3^2\right)
   +\gamma_3^2 \left(\tilde{n}^2-\tilde{\omega}^2\right)
\right.
\nonumber \\
&&
\left.
\qquad \quad
   +2 \gamma_1 \gamma_2 (n_2 n_3-\omega_2 \omega_3)
   +2 \gamma_1 \gamma_3 (n_2 \tilde{n}-\omega_2 \tilde{\omega})
   +2 \gamma_2 \gamma_3 (n_3 \tilde{n}-\omega_3 \tilde{\omega})
   \right)
   \nonumber \\
   &&
   +16 \sin (2 \theta_1) \sin^2(\xi) \cos^4(\xi) (2 \gamma_2+\gamma_3 \cos (\theta_2)) (\tilde{n} \omega_2-n_2 \tilde{\omega})
   \nonumber \\
   &&
  + \gamma_3 \sin (2 \theta_1) \sin^2(\theta_2) \cos (\theta_2) \sin^4(2 \xi) \left(\gamma_1 (n_2 n_3-\omega_2
   \omega_3)+\gamma_2 \left(n_3^2-\omega_3^2\right)+\gamma_3 (n_3 \tilde{n}-\omega_3 \tilde{\omega})\right)
   \nonumber \\
   &&
   + 8 \gamma_3 \sin (\theta_1) \cos (\theta_1) \sin^2(\theta_2) \sin^4(\xi) \cos^4(\xi) 
  \times
  \nonumber \\
  &&
  \qquad 
  \times
  \left(\gamma_3 \cos^2(\theta_1)
   \left(n_2^2-\omega_2^2\right)
   -2 \gamma_3 \cos (\theta_1) \cos (\theta_2) (n_2 n_3-\omega_2 \omega_3)+\gamma_3 \cos^2(\theta_2) \left(n_3^2-\omega_3^2\right)
   \right.
   \nonumber \\
   &&
   \left.
   \qquad \quad 
   -4 \cos (\theta_1) \left(\gamma_1 \left(n_2^2-\omega_2^2\right)+\gamma_2 (n_2 n_3-\omega_2 \omega_3)+\gamma_3 (n_2 \tilde{n}-\omega_2 \tilde{\omega})\right)
   \right)
   \nonumber \\
   &&
   + 4 \gamma_3 \sin (2 \theta_1) \sin^2(\theta_2) \sin^4(\xi) \cos^2(\xi) (n_3 \omega_2-n_2 \omega_3)
   \nonumber \\
   &&
   -16 \sin (\theta_1) \sin^2(\theta_2) \sin^4(\xi) \cos^2(\xi) (\gamma_1 (n_3 \omega_2-n_2 \omega_3)+\gamma_3 (\tilde{n} \omega_3-n_3 \tilde{\omega})).
\end{eqnarray}
\begin{eqnarray}
A_{\theta_2}^{\gamma_i} &=&
8 \sin(\theta_2) \sin^2(\xi) \left(\cos(\theta_2) (n_3-\omega_3) (n_3+\omega_3)+2 \cos^2(\xi) (n_3 \tilde{n}-\omega_3 \tilde{\omega})\right)
   \nonumber \\
   &&
   + 2 \gamma_3 \sin^2(\theta_1) \sin(2 \theta_2) \sin^2(2 \xi) (n_2 \omega_3-n_3 \omega_2)
\nonumber \\
&&
+ 8 \sin^2(\theta_1) \sin(\theta_2) \sin^2(\xi) \cos^4(\xi) 
\times
\nonumber \\
&&
\quad
\times
(2 \gamma_2 (n_3 \omega_2-n_2 \omega_3)+\gamma_3
   \cos(\theta_2) (n_3 \omega_2-n_2 \omega_3)+2 \gamma_3 (n_2 \tilde{\omega}-\tilde{n} \omega_2))
   \nonumber \\
   &&
   -32 \sin(\theta_2) \cos(\theta_2) \sin^4(\xi) \cos^2(\xi) (2 \gamma_1-\gamma_3 \cos(\theta_1)) (\tilde{n} \omega_3-n_3 \tilde{\omega})
   \nonumber \\
   &&
   + 8 \sin^2(\theta_1) \sin(\theta_2) \cos(\theta_2) \sin^4(\xi) \cos^4(\xi) 
   \times
   \nonumber \\
   &&
   \quad
   \times
\left(
\gamma_3^2 \cos^2(\theta_1) \left(n_2^2-\omega_2^2\right)-2 \gamma_3^2 \cos (\theta_1) \cos (\theta_2) (n_2
   n_3-\omega_2 \omega_3)+\gamma_3^2 \cos^2(\theta_2) \left(n_3^2-\omega_3^2\right)
   \right.
   \nonumber \\
   &&
\qquad\quad \left.
-4 \gamma_3 \cos (\theta_1) \left(\gamma_1 \left(n_2^2-\omega_2^2\right)+\gamma_2 (n_2 n_3-\omega_2
   \omega_3)+\gamma_3 (n_2 \tilde{n}-\omega_2 \tilde{\omega})\right)   
   \right.
   \nonumber \\
   &&
\left.
\qquad\quad
+4 \gamma_3 \cos (\theta_2) \left(\gamma_1 (n_2 n_3-\omega_2 \omega_3)+\gamma_2 \left(n_3^2-\omega_3^2\right)+\gamma_3 (n_3 \tilde{n}-\omega_3 \tilde{\omega})\right)
   \right.
   \nonumber \\
   &&
\left.
\qquad\quad
+ 4 \left(\gamma_1^2 \left(n_2^2-\omega_2^2\right)+2 \gamma_1 \gamma_2 (n_2 n_3-\omega_2 \omega_3)+2 \gamma_1
   \gamma_3 (n_2 \tilde{n}-\omega_2 \tilde{\omega})
   \right.
\right.
   \nonumber \\
   &&
\left.
\left.
\qquad\qquad
   +\gamma_2^2 \left(n_3^2-\omega_3^2\right)+2 \gamma_2
   \gamma_3 (n_3 \tilde{n}-\omega_3 \tilde{\omega})+\gamma_3^2 \left(\tilde{n}^2-\tilde{\omega}^2\right)\right)   
   \right).
\end{eqnarray}
\begin{eqnarray}
 B^{\gamma_i} &=& 
 64 \sin^2(\theta_2) \sin^2(\xi) \left(n_3^2-\omega_3^2\right)
 \nonumber \\
 &&
 +\gamma_3^2 \sin^2(2 \theta_1) \sin^2(\theta_2) \sin^4(2 \xi) \left(n_2^2-\omega_2^2\right)
 \nonumber \\
 &&
 + 32 \gamma_3 \sin^2(\theta_1) \sin^2(\theta_2) \sin^2(2 \xi) (n_2 \omega_3-n_3 \omega_2)
 \nonumber \\
 &&
 -8 \gamma_3 \sin (\theta_1) \sin (2 \theta_1) \sin^2(\theta_2) \sin^4(2 \xi) \left(\gamma_1 \left(n_2^2-\omega_2^2\right)+\gamma_2 (n_2 n_3-\omega_2 \omega_3)+\gamma_3 (n_2 \tilde{n}-\omega_2 \tilde{\omega})\right)
   \nonumber \\
   &&
   +64 \cos^2(\xi) \left(\sin^2(\theta_1) \left(n_2^2 -\omega_2^2\right)
   -4 \sin^2(\theta_2) \sin^4(\xi) (2 \gamma_1-\gamma_3 \cos(\theta_1)) (\tilde{n} \omega_3-n_3 \tilde{\omega})\right)
   \nonumber \\
   &&
  +  8 \gamma_3 \sin^2(\theta_1) \sin (\theta_2) \sin (2 \theta_2) \sin^4(2 \xi) \left(-\omega_3 (\gamma_1 \omega_2+\gamma_2 \omega_3+\gamma_3 \tilde{\omega})
  +\gamma_1 n_2 n_3+\gamma_2 n_3^2+\gamma_3 n_3 \tilde{n}\right)   
   \nonumber \\
   &&
   + 256 \sin^2(\theta_1) \sin^2(\xi) \cos^4(\xi) (2 \gamma_2+\gamma_3 \cos (\theta_2)) (\tilde{n} \omega_2-n_2 \tilde{\omega})
   \nonumber \\
   &&
   + \sin^2(2 \xi) \left(\gamma_3^2 \sin^2(\theta_1) \sin^2(2 \theta_2) \sin^2(2 \xi) \left(n_3^2-\omega_3^2\right)
   +64 \left(\tilde{n}^2-\tilde{\omega}^2\right)\right)
   \nonumber \\
   &&
   + 8 \sin^2(\theta_1) \sin^2(\theta_2) \sin^4(2 \xi) 
  \times
   \nonumber \\
   &&
   \times
   \left(2 \gamma_1^2 \left(n_2^2 -\omega_2^2\right)
   -\gamma_3^2 \cos(\theta_1) \cos(\theta_2) (n_2 n_3-\omega_2 \omega_3)
   \right.
   \nonumber \\
   &&
   \left.
   \qquad
   +4 \gamma_1 (\gamma_2 (n_2 n_3-\omega_2 \omega_3)+\gamma_3 (n_2 \tilde{n}-\omega_2 \tilde{\omega}))+2 \gamma_2^2 \left(n_3^2-\omega_3^2\right)
   \right.
   \nonumber \\
   &&
   \left.
   \qquad
   +4 \gamma_2 \gamma_3 (n_3 \tilde{n}-\omega_3 \tilde{\omega}) +2 \gamma_3^2 \left(\tilde{n}^2-\tilde{\omega}^2\right)\right).
\end{eqnarray}

\end{document}